\documentclass{elsart}

\begin{document}

\begin{frontmatter}

\title{Effect of Aharonov-Bohm Phase on Spin Tunneling}

\author{ChangSoo Park \thanksref{park}}
\address{Department of Physics and Astronomy, Northwestern
University, Evanston, Illinois 60202}
\author{D. K. Park}
\address{Department of Physics, Kyungnam University, Masan, 631-701,
              Korea}
\thanks[park]{Permanent address: Department of Physics,
                   Dankook University, Cheonan, 330-714, Korea.}
\begin{abstract}
The role of Aharonov-Bohm effect in quantum tunneling is examined
when a potential is defined on the $S^1$ and has $N$-fold
symmetry. We show that the low-lying energy levels split from the
$N$-fold degenerate ground state oscillate as a function of the
Aharonov-Bohm phase, from which general degeneracy conditions
depending on the magnetic flux is obtained. We apply these results
to the spin tunneling in a spin system with $N$-fold rotational
symmetry around a hard axis.\\ 03.65.Bz, 03.65.Sq, 75.45.+j,
75.10Dg.
\end{abstract}

\end{frontmatter}

\section{Introduction}
\label{intro}

Quantum tunneling in mesoscopic spin systems with large spin has
attracted much attention over a decade, because of its fundamental
and practical interests\cite{gunther}. From a fundamental point of
view, these systems reveal many interesting quantum mechanical
phenomena such as the tunneling of a large spin out of a
metastable potential minimum, termed as the macroscopic quantum
tunneling, and coherent spin tunneling between classically
degenerate potential minima, the macroscopic quantum
coherence\cite{legget}. A more surprising phenomenon in these
systems is the perfect quenching of the tunnel splitting between
two degenerate ground states. This phenomenon was predicted
theoretically\cite{loss,garg93} and observed in recent
experiment\cite{wern99} in the spin system which has two-fold
degenerate classical ground states.

The quenching of spin tunneling in a spin system is based on the
quantum interference of the topological phase, known as Berry
phase. More explicitly, consider a biaxial spin system which has
two-fold symmetry about a hard axis so that there are two
degenerate ground states along each direction of an easy axis. In
the semiclassical approach, the topological phase formed by two
symmetric instanton paths which connect the two degenerate minima
in opposite directions gives rise to a destructive interferenece,
and hence leads to a vanishing tunnel splitting for half-integer
spin (i.e., the Kramers degeneracy)\cite{loss}. When a magnetic
field is applied along the hard axis the two instanton paths are
still symmetric and wind the hard axis in opposite directions to
give the topological phase interference which depends on the
applied field. The tunnel splitting in this case oscillates as the
field varies, which is responsible for the field dependent
quenching effect\cite{garg93}.

In this letter we show that the quenching of the spin tunneling
can be understood from the viewpoint of Aharonov-Bohm(AB)
effect\cite{aha59} when the effective potential for the spin
tunneling can be considered as a two-fold symmetric potential on
$S^1$. More generally, when a potential is defined on the $S^1$
and has an $N$-fold symmetry, we investigate the effect of the
AB-phase on level splittings from the $N$-fold degenerate ground
state due to the tunneling. Using the semiclassical methods (i.e.,
the instanton approach), we develop an AB tunneling amplitude for
the tunneling of a particle in the $N$-fold symmetric potential on
the $S^1$, from which the level splittings are found to oscillate
as a function of the AB-phase. Furthermore, general degeneracy
conditions depending on the AB-phase are also found from the
oscillating level splittings. By applying these results to the
spin tunneling in a spin system with $N$-fold rotational symmetry
around the hard anisotropy axis we derive general degeneracy
conditions depending on both the magnetic field and spin. When
$N=2$ our results reproduce the same quenching conditions found in
references \cite{loss,garg93}, which confirms that the
Aharonov-Bohm effect leads to the quenching of the spin tunneling.
In the following Section, starting from a one-dimensional periodic
potential, we construct the AB tunneling amplitude in $S^1$. The
energy eigenvalues depending on the AB-phase and degeneracy
conditions will be derived from this amplitude. In Section
\ref{apspin}, we present the application of these results to the
spin tunneling problem. Finally, there will be a summary in
Section \ref{sum}.

\section{AB tunneling amplitude in $S^1$}
\label{ABtun}

Consider a particle in a one-dimensional periodic potential
$V(\phi)$ with period $2\pi$ and global minima at $\phi = 0, \pm 2
\pi, \pm 4 \pi, \cdots$, so that the Hamiltonian is
$\hat{\mathcal{H}} = \frac{p^2}{2M} + V(\phi)$, where $p =
-i\partial/\partial\phi$ is the momentum operator, and $M$ is the
particle mass. The transition amplitude for a tunneling between
two adjacent potential minima can be obtained from Euclidean time
formalism. Introducing an Euclidean time $\tau = -it$, the
Euclidean amplitude for the tunneling from $\phi=0$ at $\tau' =
-\tau/2$ to $\phi=2\pi$ at $\tau' = \tau/2$ is expressed as
\begin{equation}
{\mathcal{K}}^{PP}(2\pi, \tau) \equiv \langle 2\pi\mid
e^{-\hat{\mathcal{H}}\tau}\mid 0 \rangle^{PP} = \int
{\mathcal{D}}\phi(\tau') \exp\left[-S_E[\phi(\tau')]\right],
\label{k1}
\end{equation}
where ``$PP$'' stands for the periodic potential, and
\begin{equation}
\label{action1} S_E[\phi(\tau')] = \int_{-\tau/2}^{\tau/2}d\tau'
\left[\frac{M}{2}\left(\frac{d\phi}{d\tau'}\right)^2 + V(\phi)
\right]
\end{equation}
is the Euclidean action for the periodic potential $V(\phi)$.
Since we are interested in the tunneling between two neighboring
ground states the limit $\tau \rightarrow \infty$ is assumed.  In
the semiclassical approximation ${\mathcal{K}}^{PP}(2\pi,\tau)$
can be calculated using the dilute gas approximation and expressed
by
\begin{equation}
{\mathcal{K}}^{PP}(2\pi, \tau) = \sum_{n_1 ,n_2} \delta_{n_1 -n_2
,1} {\mathcal{G}}_{n_1 ,n_2}(2\pi, \tau), \label{KPP}
\end{equation}
where ${\mathcal{G}}_{n_1 ,n_2}(2\pi, \tau)$ is the contribution
of $n_1$ instantons and $n_2$ anti-instantons to the Euclidean
amplitude. The general form of this is given by \cite{raja82}
\begin{equation}
{\mathcal{G}}_{n_1 ,n _2}(2\pi, \tau) = \frac{e^{-(n_1 + n_2
)S_{\rm cl}}}{n_1 ! n_2 !} (D\tau)^{n_1 + n_2 } e^{-\frac{\omega
\tau }{2}}\sqrt{\frac{\omega}{\pi}}, \label{g1}
\end{equation}
where $S_{cl}$ is the Euclidean classical instanton action,
$\omega$ is equal to $\sqrt{d^2 V(\phi)/d\phi^2}$ evaluated at the
potential minima $\phi = 0$ or $2\pi$, and $D$ is some real number
which arises from the collective coordinate treatment of time
translational symmetry \cite{collect}. The explicit form of $D$,
however, is not necessary for further consideration.

Now, using the identity
\begin{equation}
\label{ide} \delta_{n_1 - n_2, N} \equiv \int_{-\pi}^{\pi}
\frac{d\vartheta}{2 \pi} e^{-i(n_1 - n_2 -N) \vartheta}
\end{equation}
with $N=1$, we can show that
\begin{equation}
\label{pp2} {\mathcal{K}}^{PP}(2\pi, \tau) = \int_{-\pi}^{\pi}
\frac{d \vartheta}{2 \pi} e^{i \vartheta}
\sqrt{\frac{\omega}{\pi}} \exp \left[-\left(\frac{\omega}{2} - 2D
e^{-S_{cl}}\cos\vartheta\right)\tau \right].
\end{equation}
Since $\phi$ runs from $-\infty$ to $\infty$ the lowest energy
levels are infinitely degenerate when the tunneling is absent.
However, when the tunneling exists, these levels will be split and
form a continuous band of energy levels. Introducing
$\hat{\mathcal{H}}\mid \psi_{\vartheta}> = E_{\vartheta}\mid
\psi_{\vartheta}>$, where $E_{\vartheta}$ is an energy level in
the lowest energy band, and $\mid\psi_{\vartheta}>$ is the
corresponding eigenfunction, the Euclidean transition amplitude
for the tunneling, in the limit $\tau \rightarrow \infty$, can
also be written as
\begin{equation}
{\mathcal{K}}^{PP}(2\pi,\tau) = \int_{-\pi}^{\pi} d\vartheta <2
\pi \mid \psi_{\vartheta} > <\psi_{\vartheta} \mid 0
> e^{-E_{\vartheta}\tau}.
\label{eigenK}
\end{equation}
Comparing this with Eq.~(\ref{pp2}) we can see that the spectrum
of the lowest energy band is
\begin{equation}
E_{\vartheta} = \frac{\omega}{2} - 2De^{-S_{\rm cl}}\cos\vartheta,
\label{band}
\end{equation}
and the wavefunction $\psi_{\vartheta}(\phi)$ satisfies the
relation
\begin{equation}
\psi_{\vartheta}(2\pi) = e^{i\vartheta}\psi_{\vartheta}(0).
\label{phase1}
\end{equation}
As we can anticipate from the periodic property of the potential
$V(\phi)$, Eqs.~(\ref{band}) and (\ref{phase1}) represent the
well-known band structure and Bloch theorem for an electron in
one-dimensional periodic lattice.

If we impose a periodic boundary condition on the wavefunction
such that
\begin{equation}
\psi_{\vartheta}(2\pi) = \psi_{\vartheta}(0) \label{bc}
\end{equation}
the above potential can be considered as a particle moving on a
circle. In this case, since $\phi$ is restricted in the range $0
\leq \phi \leq 2\pi$, the periodic potential $V(\phi)$ is reduced
to a 1-fold symmetric potential on the $S^1$ geometry. When the
periodic boundary condition (\ref{bc}) holds the phase factor
$\vartheta$ in Eq.~(\ref{phase1}) becomes zero, so that we can
insert $\delta(\vartheta)$ into the integral in Eq.~(\ref{pp2}).
The Euclidean amplitude for the tunneling in the $S^1$ geometry is
then obtained to be
\begin{equation}
\label{s1-1} {\mathcal{K}}^{S^1}(2\pi, \tau) = \frac{1}{2 \pi}
\sqrt{\frac{\omega}{\pi}} \exp \left[ - \left(\frac{\omega}{2} -
2D e^{-S_{\rm cl}}\right) \tau \right],
\end{equation}
where we have changed the superscript from ``$PP$'' to ``$S^1$''
to denote the $S^1$ geometry. Notice that, since there are no
degenerate levels in the 1-fold symmetric potential, the Euclidean
amplitude provides only the lowest energy level $E =
\frac{\omega}{2} - 2De^{-S_{\rm cl}}$ instead of giving an energy
band.

We now consider the effect of the Aharonov-Bohm phase on the
tunneling of a particle in the 1-fold symmetric potential on the
$S^1$ geometry. To this end we assume that an external magnetic
field ${\bf{H}}$ is applied along the axis of the $S^1$. The
wavefunction then acquires a phase factor $\Phi$ (i.e., the
magnetic flux) per winding due to the AB effect. In this case, the
ordinary winding number representation of the Euclidean amplitude
can be constructed by inserting $\sum_{m=-\infty}^{\infty}
\delta(\vartheta - m \Phi)$ to the integrand in Eq.~(\ref{pp2}).
In the problem of tunneling, however, we need only one phase
factor $\Phi$, which means only $m=1$ sector contributes to the
transition amplitude for the tunneling. The Euclidean amplitude
for this case, which we call AB {\it tunneling amplitude}, is then
\begin{equation}
\label{sab-1} {\mathcal{K}}_{\rm{AB}}^{S^1}(2\pi, \tau) =
\frac{e^{i\Phi}}{2\pi} \sqrt{\frac{\omega}{\pi}} \exp
\left[-\left( \frac{\omega}{2} - 2D e^{-S_{\rm cl}} \cos \Phi
\right) \tau \right],
\end{equation}
where we have inserted a subscript ``AB'' to specify the inclusion
of the Aharonove-Bohm effect. From this it can be seen that the
energy eigenvalue becomes
\begin{equation}
E(\Phi) = \frac{\omega}{2} - 2D e^{-S_{\rm cl}} \cos \Phi
\end{equation}
which is a periodic function of the flux $\Phi$ with period $\Phi
= 2\pi$. This implies that the energy level oscillates by varying
the magnetic field (i.e., the AB-phase).  For a given value of
$\Phi$ the energy level is shifted up by $2De^{-S_{\rm
cl}}(1-\cos\Phi)$, and the shift becomes zero whenever $\Phi =
2n\pi$, where $n$ is a non-negative integer.

Let us generalize the above argument to the case of $N$-fold $(N
\geq 2)$ symmetric potential $V_N(\phi)$ in the $S^1$ geometry.
The corresponding one-dimensional periodic potential has a period
of $2\pi/N$ and minima at $\phi = 0, \pm 2\pi/N, \cdots$. In this
case, the transition amplitude for the tunneling between adjacent
minima is represented by ${\mathcal{K}}^{PP}(2\pi/N,\tau)$, and
the contribution of $n_1$ instantons and $n_2$ anti-instantons is
${\mathcal{G}}_{n_1 ,n_2}(2\pi/N , \tau)$ which has the same form
as in Eq.~(\ref{g1}). Our task is to find a general AB tunneling
amplitude ${\mathcal{K}}_{\rm{AB}}^{S^1}(2\pi/N,\tau)$. Before we
do this, let us first calculate
${\mathcal{K}}_{\rm{AB}}^{S^1}(2\pi,\tau)$ to see how the
instanton and anti-instanton contribute to the AB tunneling
amplitude for the $N$-fold symmetric potential. Following the same
analysis as in the 1-fold case we can calculate this amplitude
within the dilute gas approximation. The only difference from the
previous case is to replace $\delta_{n_1 - n_2 , 1 }$ in
Eq.~(\ref{KPP}) by $\delta_{n_1 - n_2 , N}$. Using the identity in
Eq.~(\ref{ide}) and inserting $\delta(\vartheta - \Phi/N)$ into
the integrand in Eq.~(\ref{pp2}) we can obtain
\begin{eqnarray}
{\mathcal{K}}_{\rm{AB}}^{S^1}(2\pi, \tau) &=& \frac{1}{2\pi}
\sum_{n_1 ,n_2} e^{-i\frac{\Phi}{N} (n_1 - n_2
-N)}{\mathcal{G}}_{n_1 ,n_2}(\frac{2\pi}{N},\tau)\nonumber\\ &=&
\frac{e^{i\Phi}}{2\pi}\sqrt{\frac{\omega}{\pi}}\exp
\left[-\left(\frac{\omega}{2} - 2De^{S_{\rm
cl}}\cos\frac{\Phi}{N}\right)\tau\right]. \label{carry}
\end{eqnarray}
This shows that, through the tunneling from $\phi=0$ to
$\phi=2\pi$ in the $N$-fold symmetric potential, the instanton and
anti-instanton behave as if they carry phase factors
$\frac{\Phi}{N}$ and $-\frac{\Phi}{N}$, respectively, up to
overall phase factor in spite of their classical nature.

We are now in position to find the AB tunneling amplitude for the
$N$-fold symmetric potential. The observation mentioned above
makes it possible to compute this quantity without overall phase
factor which is irrelevant for the present problem, especially for
the computation of the level splitting due to the tunneling. Since
the instanton and anti-instanton carry their own phases, the
multi-instanton contribution to the AB tunneling amplitude for the
$N$-fold symmetric potential can be described by
\begin{equation}
\label{multi}
{\mathcal{K}}_{\mathrm{AB}}^{S^1}\left(\frac{2\pi}{N},\tau\right)
= \sum_{n_1} \sum_{n_2} P_{n_1 ,n_2}^N
{\mathcal{G}}_{n_1,n_2}\left(\frac{2\pi}{N}, \tau\right) e^{-i
\frac{\Phi}{N} (n_1 - n_2 )}
\end{equation}
where $P_{n_1, n_2}^N$ represents all possible sequence of
instantons and anti-instantons. For $N \geq 2$ this can be
expressed as
\begin{equation}
P_{n_1 ,n_2}^N = \sum_{n, m =0}^{\infty} \left[ \delta_{n_1 ,
Nm}\delta_{n_2 , Nn+N-1} + \delta_{n_1 ,Nm+1}\delta_{n_2 , Nn} +
\sum_{\eta \geq 2}^{N-1} \delta_{n_1 , Nm+\eta} \delta_{n_2 ,
Nn+\eta-1} \right],
\end{equation}
where the summation over $\eta$ exists only when $N \geq 3$. Using
this relation and the expression of
${\mathcal{G}}_{n_1,n_2}(2\pi/N, \tau)$ in Eq.~(\ref{g1}) we can
write the Eq.~(\ref{multi}) as
\begin{equation}
{\mathcal{K}}_{\mathrm{AB}}^{S^1}\left(\frac{2\pi}{N},\tau\right)
= \sqrt{\frac{\omega}{\pi}} e^{-\frac{\omega}{2}\tau} \left[
f_{N,0}(x_1 ) f_{N,N-1}(x_2 ) + f_{N,1}(x_1 )f_{N,0}(x_2) +
\sum_{\eta \geq 2}^{N-1} f_{N, \eta}(x_1) f_{N, \eta-1}(x_2 )
\right], \label{KN}
\end{equation}
where
\begin{equation}
x_1 = x_2^* \equiv D e^{-S_{\mathrm{cl}}} \left( \cos
\frac{\Phi}{N} + i \sin \frac{\Phi}{N} \right)\tau,
\end{equation}
and
\begin{equation}
f_{N,\sigma}(x) = \sum_{l=0}^{\infty}
\frac{x^{Nl+\sigma}}{(Nl+\sigma)!}
\end{equation}
which satisfies following relations;
\begin{equation}
f_{N,\sigma}(x) = \frac{d^{N-\sigma}f_{N,0}(x)}{dx^{N-\sigma}},
\hspace{1cm} \frac{d^N f_{N,0}(x)}{dx^N} = f_{N,0}(x).
\end{equation}
The boundary conditions for the differential equations are given
by
\begin{equation}
f_{N,0}(0) = 1,\hspace{1cm} f_{N, \sigma \geq 1}(0) = 0.
\end{equation}
Using these boundary conditions we can solve the differential
equation and find
\begin{equation}
f_{N,0}(x) = \cases{\frac{2}{N}\left[ \cosh x +
\sum\limits_{r=1}^{\frac{N}{2}-1} \cos \left( x \sin \frac{2\pi
r}{N} \right)\exp \left( x\cos \frac{2\pi r}{N}\right) \right],&if
$N$=even;\cr\noalign{\vskip5pt} \frac{1}{N} \left[ e^x +
2\sum\limits_{r=1}^{\frac{N-1}{2}} \cos \left( x \sin \frac{2\pi
r}{N} \right)\exp \left( x\cos \frac{2\pi r}{N} \right)
\right],&if $N$=odd. \cr}
\end{equation}
Substituting these solutions into the Eq.~(\ref{KN}) and
performing the summation over $\eta$ for each case of $N$ we
obtain
\begin{equation}
{\mathcal{K}}_{\mathrm{AB}}^{S^1}\left(\frac{2\pi}{N},\tau\right)
= \frac{1}{N} \sqrt{\frac{\omega}{\pi}}\sum_{k} \exp \left[
-\left(\frac{\omega}{2} - 2D e^{-S_{\rm{cl}}} \cos \frac{\Phi -
2\pi k}{N} \right)\tau - i \frac{2\pi k}{N} \right].
\end{equation}
Thus, the energy eigenvalues due to the tunneling in the $N$-fold
symmetric potential on the $S^1$-geometry are
\begin{equation}
E_{Nk}(\Phi) = \frac{\omega}{2} - 2D e^{-S_{\mathrm{cl}}} \cos
\frac{\Phi -2\pi k}{N}, \label{ENK}
\end{equation}
where $k$ is an integer allowed in the range
\begin{equation}
-\frac{N}{2} < k \leq \frac{N}{2}. \label{kcond}
\end{equation}

Unlike the 1-fold symmetric potential, there are now $N$
eigenvalues for a given $N$, of which all depend on the AB phase
$\Phi$. The reason for the existence of the several eigenvalues is
obvious from the fact that the potential $V_N(\phi)$ has $N$
minima within $2\pi$, and hence there are $N$-fold degenerate
lowest levels which are split by the tunneling. Note, however,
that, since each level oscillates as a function of the AB-phase,
the levels may intersect each other to become degenerate states at
certain values of $\Phi$. To see how this happens we consider the
difference between any two levels corresponding to $k$ and $k'$
$(k \neq k')$, respectively:
\begin{eqnarray}
\Delta E_N &=&  E_{Nk}(\Phi) - E_{Nk'}(\Phi)  \nonumber \\ &=&
4De^{-S_{\rm cl}} \sin \frac{\Phi - (k+k')\pi}{N}
\sin\frac{\pi(k-k')}{N}. \label{deltEN}
\end{eqnarray}
From this and the Eq.~(\ref{kcond}) we observe following general
features for the degeneracy depending on $\Phi$:
\begin{eqnarray}
\begin{array}{lll} \left. \begin{array}{l} \frac{N}{2}~~ {\rm double\ degeneracies\ at}~
\Phi = (2n + 1)\pi \\ \frac{N}{2} - 1 ~~{\rm double\ degeneracies\
at}~ \Phi = 2n\pi \end{array} \right\}& &\mbox{when $N$ = even};
\label{dege1}\\ \frac{N-1}{2}~~ {\rm double\ degeneracies\ at}~~
\Phi = n\pi,& &\mbox{when $N$ = odd},
\end{array}
\end{eqnarray}
where $n$ is a non-negative integer. Note that, in all cases, the
degree of the degeneracy is at most double. This implies that a
complete suppression of the tunnel splitting (i.e., the quenching)
is possible only when $N=2$. From above conditions, this quenching
effect occurs whenever the AB-phase becomes an odd-integer
multiple of $\pi$.

\section{Application to  spin tunneling}
\label{apspin}

As an application of above results, let us now consider the spin
tunneling in a spin system which has an $N$-fold symmetry about a
hard axis. Especially, we will look into the cases of $N = 2, 3,
4, 6$ which can be realized in orthorhombic, trigonal, tetragonal,
and hexagonal structures, respectively. We choose $z$ as the hard
axis so that $xy$ is the easy plane. We also assume that an
external magnetic field is applied along the hard axis. With this
choice the spin Hamiltonian which displays the $N$-fold symmetry
can be written as
\begin{equation}
{\hat{\mathcal{H}}}_N = A S_z^2 - C(S_{+}^N + S_{-}^N ) - g\mu_B
H_z S_z, \label{hamil}
\end{equation}
where $N = 2, 3, 4, 6$, $g=2$, $\mu_B$ is the Bohr magneton, and $
A > C > 0$. For simplicity, we have neglected the higher order
uniaxial anisotropy terms $S_z^4$ and $S_z^6$. To study the spin
tunneling within the semiclassical methods we use the spin
coherent state path integral approach.

In the spin coherent state representation the anisotropy energy
corresponding to the Hamiltonian (\ref{hamil}) is given by
\begin{eqnarray}
{\mathcal{E}}_N(\theta, \phi) &=& <{\bf{\Omega}}\mid
\hat{\mathcal{H}}_N \mid {\bf{\Omega}}> \nonumber\\ &=& AS^2
\left(\cos^2\theta - \lambda S^{N-2}\sin^N\theta\cos N\phi -
h\cos\theta \right), \label{clashamil}
\end{eqnarray}
where $\mid{\bf{\Omega}}> = \mid \theta, \phi>$ is the spin
coherent state with $\theta$, $\phi$ being the polar and azimuthal
angles, respectively,  $\lambda \equiv 2C/A$, $h \equiv g\mu_B H_z
/AS$, and $S$ is the spin which is assumed to be large so that the
semiclassical methods can be applied. Since we have chosen the $z$
axis as the hard axis $\lambda S^{N-2} < 1$ should also be
assumed. The energy ${\mathcal{E}}_N(\theta,\phi)$ exhibits
$N$-fold degenerate classical minima at $\theta = \theta_0$, $\phi
= 0, 2\pi/N, \cdot\cdot\cdot, 2(N-1)\pi/N$, where $\theta_0 =
\pi/2$ for $h = 0$, and decreases smoothly to $0$ as $h$ is
increased. Since $\phi$ is defined in the range $0 \leq \phi \leq
2\pi$, the spin tunneling from $(\theta_0, 0)$ to $(\theta_0,
2\pi/N)$ is essentially same as the tunneling in the $N$-fold
symmetric potential on the $S^1$ geometry.

To see this explicitly we write the AB tunneling amplitude for the
spin tunneling in the spin coherent state representation:
\begin{eqnarray}
{\mathcal{K}}_{\rm AB}^{S^1}\left(\frac{2\pi}{N},\tau\right) &=&
<{\bf{\Omega}}_f\mid
e^{-\hat{\mathcal{H}}\tau}\mid{\bf{\Omega}}_i>\nonumber\\ &=&\int
{\mathcal{D}}[\phi(\tau')]{\mathcal{D}}[\cos\theta(\tau')]
e^{-S_E[\theta(\tau'),\phi(\tau')]}, \label{spinK}
\end{eqnarray}
where
\begin{equation}
S_E[\theta(\tau'),\phi(\tau')]=\int_{-\tau/2}^{\tau/2}[{\mathcal{E}}_N(\theta,
\phi) - i S{\dot{\phi}}(\cos \theta - 1)]d \tau' \label{spinact}
\end{equation}
is the Euclidean action including the Berry-phase term. Here, the
boundary conditions are $\mid{\bf{\Omega}}(-\frac{\tau}{2})> =
\mid{\bf{\Omega}}_i> = \mid\theta_0, 0>$,
$\mid{\bf{\Omega}}(\frac{\tau}{2})> = \mid{\bf{\Omega}}_f> =
\mid\theta_0, 2\pi/N> $. The analogy of the spin tunneling with
the tunneling in $S^1$ geometry can be revealed by making the
action (\ref{spinact}) similar to Eq.~(\ref{action1}), which can
be done by performing the integral over $\cos\theta$ in
Eq.~(\ref{spinK}). Note, however, that the integral is not
Gaussian for $N \geq 3$. Since we are interested in the general
feature of the effect of the AB phase on the spin tunneling, we
look for an analytical result. To do this, we assume that
$\lambda$ and $h$ are small enough so that an approximation
\begin{equation}
\sin^N\theta \approx 1 -\frac{N}{2}\cos^2\theta \label{approx}
\end{equation}
can hold for $N \geq 3$. By using this approximation and
performing the $\theta$ integral we obtain an effective Euclidean
action
\begin{equation}
S_{\rm eff} = \int_{-\tau/2}^{\tau/2} d\tau
\left[\frac{M_N(\phi)}{2}{\dot{\phi}}^2 + V_N(\phi) \right] +
i\int_{-\tau/2}^{\tau/2}d\tau {\mathcal{A}}_N(\phi){\dot{\phi}},
\end{equation}
where
\begin{eqnarray}
M_N(\phi) &=& \frac{1}{2A\left(1 + \frac{\lambda N S^{N-2}}{2}\cos
N\phi \right)},\nonumber\\ \mathcal{A}_N(\phi) &=& S\left[1 -
\frac{h}{2\left(1 + \frac{\lambda N S^{N-2}}{2}\cos
N\phi\right)}\right],\\ V_N(\phi) &=& -AS^2 \left[\lambda
S^{N-2}\cos N\phi + \frac{h^2}{4\left(1 + \frac{\lambda N
S^{N-2}}{2}\cos N\phi\right)}\right].\nonumber
\end{eqnarray}
Note that the action is complex. Comparing the real part with
Eq.~(\ref{action1}), $M_N(\phi)$ can be understood as the position
dependent mass, and hence $V_N(\phi)$ become the potential that
has an $N$-fold symmetry and minima at $\phi = 0, 2\pi/N, \cdot
\cdot, 2(N-1)\pi/2$ on the $S^1$ because $\phi$ is restricted in
the range $0 \leq \phi \leq 2\pi$. The imaginary part contributes
to the AB tunneling amplitude (\ref{spinK}) as a phase factor
which plays the same role as the phases carried by the instanton
and anti-instanton: For the tunneling from $\phi = 0$ to $\phi =
2\pi/N$, the instanton picks up a phase factor $\Phi_N =
\int_0^{2\pi/N}d\phi{\mathcal A}_N(\phi)$, while the anti-istanton
acquires a phase $-\Phi_N$. When these two paths interfere each
other and make a closed loop around the hard axis along which the
magnetic field is applied, the integral in the imaginary part
become the AB-phase. From this notion the AB-phase can be
calculated by integrating the imaginary part along the $S^1$;
\begin{equation}
\Phi = \int_0^{2\pi} {\mathcal{A}}_N(\phi) d\phi = 2\pi S \left[ 1
- \frac{h}{\sqrt{4 - \lambda^2 N^2 S^{2(N-2)}}}\right],
\label{phiN}
\end{equation}
where $N = 2, 3, 4, 6$. As we have mentioned below
Eq.~(\ref{carry}), it can be seen that the phase carried by the
instanton $\Phi_N$ is equivalent to $\Phi/N$.

Substituting this into Eqs.~(\ref{ENK}) and (\ref{deltEN}) we
observe several interesting features for the level splittings due
to the spin tunneling in the $N$-fold symmetric potential. First,
we see that the energy levels split from the ground state
oscillate as the field varies, and thus intersect each other at
certain values of $h$ as expected. From the degeneracy conditions
in Eq.~(\ref{dege1}), we note that not all levels are
simultaneously degenerate in the cases of $N = 3, 4, 6$ although
there can be more than one double degeneracy for a given value of
$\Phi$. The $N=2$ case, which corresponds to the 2-fold symmetric
potential on the $S^1$ geometry, is more interesting. In this case
we do not need the approximation (\ref{approx}) to calculate the
$\theta$ integral in Eq.~({\ref{spinK}) and are able to find exact
condition for the degeneracy. Using the condition (\ref{dege1})
and Eq.~(\ref{phiN}) with $N=2$ we find that the two ground states
are degenerate whenever
\begin{equation}
h = \frac{2\sqrt{1-\lambda^2}}{S}\left(S - n -\frac{1}{2}\right)
\end{equation}
which coincides with the previously found result\cite{garg93}
noticing that $A+2C = k_1$ and $4C = k_2$. Thus, the field
dependent quenching of the spin tunneling in a 2-fold symmetric
spin system can be understood from the viewpoint of Ahronov-Bohm
effect.

Second, since we included the Berry phase term in
Eq.~(\ref{spinact}) the AB-phase in Eq.~(\ref{phiN}) provides more
information about the degeneracy. To see this we consider the
field free case, i.e., $h=0$. Then, the AB-phase becomes $\Phi =
2\pi S$. Using this and Eq.~(\ref{deltEN}) we find following
degeneracy conditions depending on the spin:
\begin{eqnarray*}
\begin{array}{llll} \left.\begin{array}{l}
\frac{N}{2}~~\mbox{double degeneracies for half-integer}~S\\
\frac{N}{2} - 1~~ \mbox{double degeneracies for integer}~ S
\end{array} \right\}& & &\mbox{when $N$ = even};\\ \frac{N-1}{2}~~
\mbox{double degeneracies for half-integer and integer}~ S, & & &
\mbox{when $N$ = odd}.
\end{array}
\end{eqnarray*}
Notice that, for $N \geq 3$, levels can be doubly degenerate for
both half-integer and integer spin and the number of double
degeneracies increases as the symmetry increases. When $N = 2$,
the above results show that there exists one double degeneracy for
half-integer spin. This means that, since there are two degenerate
ground states in this case, the level splitting due to the spin
tunneling disappears for half-integer spin. Thus, the spin parity
effect (i.e., the Kramers degeneracy) discussed in reference
\cite{loss} can be explained by the argument based on the
Aharonov-Bohm effect.

\section{Summary}
\label{sum}

We have studied the Aharonov-Bohm effect on the tunneling in an
$N$-fold symmetric potential on the $S^1$ geometry. Using the
semiclassical methods we found that the low-lying levels split
from the $N$-fold degenerate ground state oscillate as a function
of the Aharonov-Bohm phase, from which general degenracy
conditions depending on the phase are also obtained. By applying
these observations to the spin tunneling in the spin system which
has an $N$-fold rotational symmetry around the hard axis we found
oscillating tunnel splittings with varying magnetic field applied
along the hard axis and degeneracy conditions which are dependent
on the spin. From this, the quenching of the spin tunneling in the
2-fold symmetric spin system can be interpreted as the
Aharonov-Bohm effect.

\ack This work is supported by Dankook University through the
University Research Fund.

\end{document}